\documentclass[twocolumn,showpacs,floatfix,superscriptaddress,amsmath,amssymb,prl]{revtex4}
\usepackage{bbm}
\usepackage{mathrsfs}
\usepackage{txfonts}
\usepackage{amssymb}
\usepackage[dvips]{graphicx}
\usepackage{hyperref}
\usepackage{ulem}
 \usepackage{overpic}
 \usepackage{psfrag}
 \newcommand{\be}{\begin{equation}}
\newcommand{\ee}{\end{equation}}

\begin{document}
\title{An efficient computation of geometric entanglement for two-dimensional quantum lattice systems}

\author{ Hong-Lei Wang, Qian-Qian Shi, Sheng-Hao Li, and Huan-Qiang Zhou }
\affiliation{Centre for Modern Physics and Department of Physics,
Chongqing University, Chongqing 400044, The People's Republic of
China}

\begin{abstract}
The geometric entanglement per lattice site, as a holistic measure
of the multipartite entanglement, serves as a universal marker to
detect quantum phase transitions in quantum many-body systems.
However, it is very difficult to compute the geometric entanglement
due to the fact that it involves a complicated optimization over all
the possible separable states. In this paper, we propose a
systematic method to efficiently compute the geometric entanglement
per lattice site for quantum many-body lattice systems in two
spatial dimensions in the context of a newly-developed tensor
network algorithm based on an infinite projected entangled pair
state representation.  It is tested for quantum Ising model in a
transverse magnetic field and anisotropic spin $1/2$
anti-ferromagnetic XYX model in an external magnetic field on an
infinite-size square lattice. In addition, the geometric
entanglement per lattice site is able to detect the so-called
factorizing field. Our results are in a quantitative agreement with
Quantum Monte Carlo simulations.
\end{abstract}


\maketitle

{\it Introduction.} The geometrical entanglement is a measure of the
multipartite entanglement present in a quantum state wave function.
It quantifies the distance between a given quantum state wave
function and the closest separable state~\cite{hba,tcw}. Remarkably,
it serves as an alternative marker~\cite{tcw,rt,rtt} to locate
critical points for quantum many-body lattice systems undergoing
quantum phase transitions (QPTs)~\cite{sachdev,wen}. In addition, an
intriguing connection to both the Renormalization Group and
Conformal Field Theory has been unveiled for quantum many-body
lattice systems in one spatial dimension~\cite{abot,qqs,jean}.
Furthermore, recent numerical simulations~\cite{qqs} established a
universal finite-size correction to the geometric entanglement for
the critical XXZ and transverse quantum Ising chains, which in turn
is related with the celebrated Affleck-Ludwig $g$
factors~\cite{jean,hu}. Therefore, the geometrical entanglement
offers a powerful tool to investigate quantum criticality in quantum
many-body lattice systems. However, almost all studies have been
restricted to quantum many-body lattice systems in one spatial
dimension, with an exception~\cite{hl}, in which the geometrical
entanglement is exploited to study QPTs for quantum many-body
lattice systems in two spatial dimensions. This is mainly due to the
fact that it is very difficult to compute the geometrical
entanglement, because it involves a complicated optimization over
all the possible separable states.

On the other hand, significant progress has been made to develop
efficient numerical algorithms to simulate quantum many-body lattice
systems in the context of the tensor network
representations~\cite{verstraete,vidal,gv2d,bvt,rosgv,pwe,sz,hco,jwx,wks}.
The algorithms have been successfully exploited to compute the
ground-state fidelity per lattice site~\cite{zhou,zov,jhz,whl,lsh},
a universal marker to detect QPTs, for quantum many-body lattice
systems. Indeed, the ground-state fidelity per lattice site is
closely related to the geometrical entanglement. Therefore, it is
natural to expect that there should be an efficient way to compute
the  geometrical entanglement in the context of the tensor network
algorithms. Actually, this has been achieved for quantum many-body
lattice systems with the periodic boundary conditions in one spatial
dimension~\cite{hu} in the context of the matrix product state
representation.

In the present work,  we propose a systematic method to efficiently
compute the geometric entanglement per lattice site for quantum
many-body lattice systems in two spatial dimensions in the context
of a newly-developed tensor network algorithm based on an infinite
projected entangled pair state (iPEPS) representation.  It is
exploited to evaluate the geometric entanglement per lattice site
for quantum Ising model in a transverse magnetic field and
anisotropic spin $1/2$ anti-ferromagnetic XYX model in an external
magnetic field on an infinite-size square lattice, which enables us
to identify their critical points. In addition, the geometric
entanglement per lattice site is able to detect the so-called
factorizing field. Our results are in a quantitative agreement with
Quantum Monte Carlo simulations.

{\it The geometric entanglement per lattice site.} For a pure
quantum state $|\psi\rangle$ with $N$ parties, the geometric
entanglement, as a global measure of the multipartite entanglement,
quantifies the deviation from the closest separable state
$|\phi\rangle$. Mathematically, the geometric
entanglement~\cite{qqs,jean} $ E(|\psi\rangle)$ for an $N$-partite
quantum state $|\psi\rangle$ is expressed as:
\begin{equation}
 E(|\psi\rangle)=-\log_2{\Lambda_{{\rm max}}^{2}}, 
\end{equation}
where $\Lambda_{\rm max}$ is the maximum fidelity between $|\psi\rangle$ and all the possible
separable and normalized state $|\phi\rangle$, with
\begin{equation}
 \Lambda_{{\rm max} }={\rm \max_{|\phi\rangle }} \; |\langle\psi|\phi\rangle|.
\end{equation}
Then, the geometric entanglement per party ${\mathcal
E}_{N}(|\psi\rangle)$ is defined as:
\begin{equation}
 {\mathcal E}_{N}(|\psi\rangle)=N^{-1}E(|\psi\rangle).
\end{equation}
It corresponds to the maximum fidelity per party $\lambda^{{\rm
max}}_{N}$, where
\begin{subequations}
\begin{equation}
 \lambda^{{\rm max}}_{N}=\sqrt[N]{{\Lambda_{{\rm max}}}},
\end{equation}
or, equivalently,
\begin{equation}
{\mathcal E}_{N}(|\psi\rangle)=-\log_2{{(\lambda^{{\rm
max}}_{N})}^{2}}.
\end{equation}
\end{subequations}

For our purpose, we shall consider a quantum many-body system on an
infinite-size lattice in two spatial dimensions, which undergoes a
QPT at a critical point in the thermodynamic limit. In this
situation, each lattice site constitutes a party, thus the geometric
entanglement per party  becomes the geometric entanglement per
lattice site, which is well defined even in the thermodynamic limit
($N \rightarrow \infty$), since the contribution to fidelity from
each party (site) is multiplicative.

\begin{figure}
\includegraphics[width=0.48\textwidth]{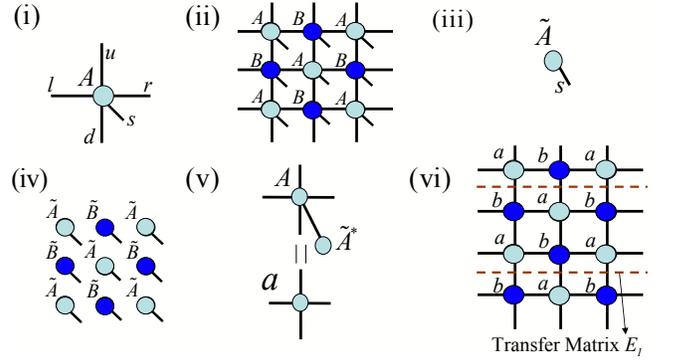}
\caption{(color online) (i) A five-index tensor $A^s_{lrud}$ labeled
by one physical index $s$ and four bond indices $l$, $r$, $u$ and
$d$. (ii) The iPEPS representation of a wave function in
two-dimensional square lattice. Copies of the tensors $A^s_{lrud}$
and $B^s_{lrud}$ are connected through four types of bonds. (iii) A
one-index tensor $\tilde{A}^{s}$ labeled by one physical index $s$.
(iv) The iPEPS representation of a separable state in
two-dimensional square lattice. (v) A reduced four-index tensor
$a_{lrud}$ from a five-index tensor $A^s_{lrud}$ and a one-index
$\tilde{A}^{s*}$.  (vi) The tensor network representation for the
fidelity between a quantum wave function (describe by $A^s_{lrud}$
and $B^s_{lrud}$) and a separable state (described by $\tilde{A}^s$
and $\tilde{B}^s$), consisting of the reduced tensors $a_{lrud}$ and
$b_{lrud}$. } \label{FIG1}
\end{figure}

{\it The infinite projected entangled pair state algorithm.} Our aim
is to compute the geometric entanglement per lattice site for a
quantum many-body lattice system on an infinite-size square lattice
in the context of the iPEPS algorithm~\cite{gv2d}.  Suppose we
consider a system characterized by a translation-invariant
Hamiltonian $H$ with the nearest-neighbor interactions:
$H=\sum_{<ij>} h_{<ij>}$, with $h_{<ij>}$ being the Hamiltonian
density. Assume that a quantum wave function $|\psi\rangle$ is
translation-invariant under two-site shifts, then one only needs two
five-index tensors $A^s_{lrud}$ and $B^s_{lrud}$ to express the
iPEPS representation. Here, each tensor is labeled by one physical
index $s$ and four bond indices $l$, $r$, $u$ and $d$, as shown in
Fig.\ref{FIG1}(i). Note that the physical index $s$ runs over
$1,\cdots,\mathbbm{d}$, and each bond index takes $1,\cdots,D$, with
$\mathbbm{d}$ being the physical dimension, and $D$  the bond
dimension. Therefore, it is convenient to choose a $2\times 2$
plaquette as the unit cell (cf. Fig.\ref{FIG1}(ii)). The
ground-state wave function is well approximated by
$|\psi_{\tau}\rangle$, which is obtained by performing an imaginary
time evolution~\cite{gv2d} from an initial state $|\psi_{0}\rangle$,
with
$|\psi_{\tau}\rangle=e^{-H\tau}|\psi_{0}\rangle/||e^{-H\tau}|\psi_{0}\rangle||$~\cite{gv2d},
as long as $\tau$ is large enough.

A key ingredient of the iPEPS algorithm is to take advantage of the
Trotter-Suzuki decomposition that allows to reduce the (imaginary)
time evolution operator $e^{-H \delta\tau}$ over a time slice
$\delta \tau$ into the product of a series of two-site operators,
where the imaginary time interval $\tau$ is divided into $M$ slices:
$\tau = M \delta \tau$. Therefore, the original global optimization
problem becomes a local two-site optimization problem. With an
efficient contraction scheme available to compute the effective
environment for a pair of the tensors $A^s_{lrud}$ and $B^s_{lrud}$
~\cite{gv2d}, one is able to update the tensors $A^s_{lrud}$ and
$B^s_{lrud}$. Performing this procedure until the energy per lattice
site converges, the ground-state wave function is produced in the
iPEPS representation.

\begin{figure}
\includegraphics[width=0.48\textwidth]{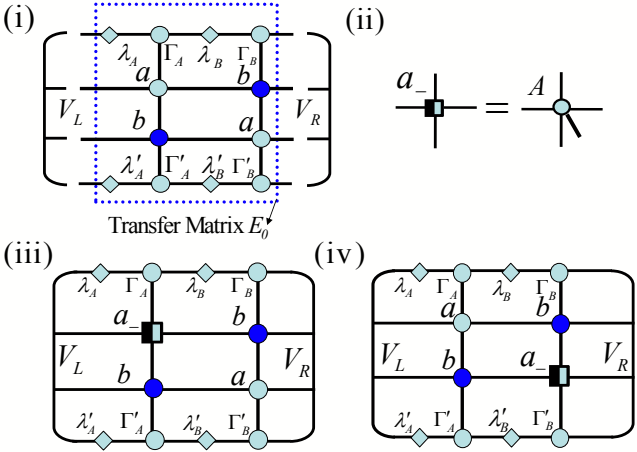}
\caption{(color online) The gradient of the fidelity between a given
ground-state wave function $|\psi \rangle$ and a separable state
$|\phi \rangle$ in the iPEPS representation. (i): The
zero-dimensional transfer matrix $E_0$ and its dominant eigenvectors
$V_{L}$ and $V_{R}$. Here, the infinite matrix product state
representation of the dominant eigenvectors for  the one-dimensional
transfer matrix $E_1$ follows from Ref.~\cite{rosgv}, and $V_{L}$
and $V_{R}$ may be evaluated from the Lanczos method. The
contraction of the entire tensor network is the dominant eigenvalue
$\eta_{\langle\phi|\psi\rangle}$ of the zero-dimensional transfer
matrix $E_0$ for $\langle\phi|\psi\rangle$. (ii): A half-filled
square denotes $a_{-}$, the derivative of the four-index tensor
$a_{lrud}$ with respect to $\tilde{A}^{s*}$, which is nothing but
the five-index tensor $A^s_{lrud}$. Similarly, we may define
$b_{-}$, the derivative of the four-index tensor $b_{lrud}$ with
respect to $\tilde{B}^{s*}$. (iii) and (iv): The pictorial
representation of the contributions to the derivative of
$\eta_{\langle\phi|\psi\rangle}$ with respect to $\tilde{A}^{s*}$,
with different relative positions between filled circles and
half-filled squares. } \label{FIG2}
\end{figure}

{\it Efficient computation of the geometric entanglement in the
iPEPS representation.} Once the iPEPS representation for the
ground-state wave function is generated, we are ready to evaluate
the geometric entanglement per lattice site. First, we need to
compute the fidelity between the ground-state wave function and a
separable state. The latter is represented in terms of one-index
tensors $\tilde A^s$ and $\tilde B^s$. To this end, we form a
reduced four-index tensor $a_{lrud}$ from the five-index tensor
$A^s_{lrud}$ and a one-index tensor $\tilde A ^s$, as depicted in
Fig.~\ref{FIG1}\;(iii). As such, the fidelity is represented as a
tensor network in terms of the reduced tensors $a_{lrud}$ and
$b_{lrud}$ (cf. Fig.~\ref{FIG1}\;(iv)). The tensor network may be
contracted as follows. First, form the one-dimensional transfer
matrix $E_1$, consisting of two consecutive rows of the tensors in
the checkerboard tensor network. This is highlighted in
Fig.~\ref{FIG1}\;(vi) with two dash lines. Second, compute the
dominant eigenvectors of the transfer matrix $E_1$, corresponding to
the dominant eigenvalue. This can be done, following a procedure
described in Ref.~\cite{rosgv}. Here, the dominant eigenvectors are
represented in the infinite matrix product states.  Third, choose
the zero-dimensional transfer matrix $E_0$, and compute its dominant
left and right eigenvectors, $V_{L}$ and $V_{R}$. This may be
achieved by means of the Lanczos method. In addition, one also needs
to compute the norms of the ground-state wave function
$|\psi\rangle$ and a separable state $|\phi\rangle$ from their iPEPS
representations. Putting everything together, we are able to get the
fidelity per unit cell between the ground state $|\psi\rangle$ and a
separable state $|\phi\rangle$:
\begin{equation}
  \lambda=\frac{|\eta_{\langle\phi|\psi\rangle}|}
  {\sqrt{\eta_{\langle\psi|\psi\rangle}\eta_{\langle\phi|\phi\rangle}}},
\end{equation}
where  $\eta_{\langle\phi|\psi\rangle}$, $\eta_{\langle\phi|\phi\rangle}$ and
$\eta_{\langle\phi|\phi\rangle}$ are, respectively, the dominant
eigenvalue of the zero-dimensional transfer matrix $E_0$ for the
iPEPS representation of $\langle\phi|\psi\rangle$, $\langle\psi|\psi\rangle$ and
$\langle\phi|\phi\rangle$.
Then we proceed to compute the geometric entanglement per lattice
site, which involves the optimization over all the separable states.
For our purpose, we define $F=\lambda^{2}$. The optimization amounts
to computing the logarithmic derivative of $F$ with respect to
$\tilde{A}^{*}$, which is expressed as
\begin{equation}
G \equiv \frac{\partial \ln F}{\partial
\tilde{A}^{*}}=\frac{1}{\eta_{\langle\phi|\psi\rangle}}\frac{\partial\eta_{\langle\phi|\psi\rangle}}{\partial
\tilde{A}^{*}}-\frac{1}{\eta_{\langle\phi|\phi\rangle}}\frac{\partial\eta_{\langle\phi|\phi\rangle}}{\partial
\tilde{A}^{*}}.
\end{equation}
Therefore, the problem reduces to the computation of $G$ in the
context of the tensor network representation. First, note that a
pictorial representation of the derivative $\partial a_{lrud}/
\partial \tilde{A}^{s*}$ of the four-index tensor $a_{lrud}$ with respect to
$\tilde{A}^{s*}$ is shown in Fig.~\ref{FIG2}\;(ii), which is nothing
but the five-index tensor $A^s_{lrud}$. Similarly, we may define the
derivative of the four-index tensor $b_{lrud}$ with respect to
$\tilde{B}^{s*}$. Then, we are able to represent the contributions
to the derivative of $\eta_{\langle\phi|\psi\rangle}$ with respect
to $\tilde{A}^{s*}$ in Fig.~\ref{FIG2}\;(iii) and (iv). In our
scheme, we update the real and imaginary parts of $\tilde{A}^s$
separately:
\begin{subequations}
\begin{equation}
\Re(\tilde{A}^s)=\Re(\tilde{A}^s)+\delta \Re(G)^{s},
\end{equation}
and
\begin{equation}
\Im(\tilde{A}^s)=\Im(\tilde{A}^s)+\delta \Im(G)^{s}.
\end{equation}
\end{subequations}
Here, $\delta\in[0,1)$ is the step size in the parameter space,
which is tuned to be decreasing during the optimization process. In
addition, we have normalized the real and imaginary parts of the
gradient $G$ so that their respective largest entries are unity. The
procedure to update the tensor $\tilde{B}^s$ is the same. If the
fidelity per unit cell converges, then the closest separable state
$|\phi\rangle$ is achieved, thus the geometric entanglement per
lattice site for the ground-state wave function $|\psi\rangle$
follows.

{\it The models.} As a test, we simulate two typical quantum
many-body lattice systems on an infinite-size square lattice in two
spatial dimensions. The first model is quantum Ising model in a
transverse magnetic field. The Hamiltonian takes the form,
\begin{equation}
  H=-\sum_{<i,j>}S^{[i]}_{x}S^{[j]}_{x}-h \sum_{i} S^{[i]}_{z}, 
\end{equation}
where $S^{[j]}_{\alpha} (\alpha=x,z)$ are the spin-$1/2$ Pauli
operators at lattice site $j$, $h$ is a transverse magnetic field,
and $< i, j >$ runs over all the possible pairs of the nearest
neighbors on a square lattice. Quantum Monte Carlo simulation
predicts a critical point at $h_{\rm QMC} \sim 3.044$~\cite{hwj}.

The second model is quantum anti-ferromagnetic XYX model in an
external magnetic field. The Hamiltonian is written as,
\begin{equation}
  H=J\sum_{<i,j>} \left(S^{[i]}_{x}S^{[j]}_{x}+\Delta_{y}
  S^{[i]}_{y}S^{[j]}_{y}+S^{[i]}_{z}S^{[j]}_{z}\right)+h \sum_{i} S^{[i]}_{z}, 
\end{equation}
where $S^{[j]}_{\alpha} (\alpha=x,y,z)$ are the spin-$1/2$ Pauli
operators at lattice site $j$, $J > 0$ is the exchange coupling, $<
i, j >$ runs over all the possible pairs of the nearest neighbors on
a square lattice, and $h$ is an external magnetic field. Note that
$\Delta_{y}<1$ and $\Delta_{y}>1$ correspond to the easy-plane and
easy-axis behaviors, respectively. Here, we focus on $\Delta_{y} =
0.25$. In this case, the model is known to undergo a continuous QPT
in the same universality class as the transverse Ising
model~\cite{tros}.

\begin{figure}
\includegraphics[width=0.42\textwidth]{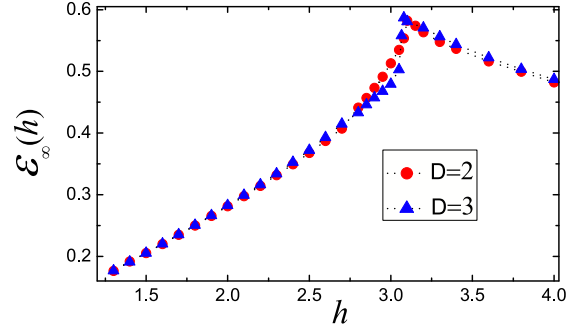}
\caption{(color online) The geometric entanglement per lattice site
${\mathcal E}_{\infty}(h)$ as a function of the transverse magnetic
field strength $h$ for quantum transverse Ising model on an
infinite-size square lattice in two spatial dimensions. The data are
presented for both the truncation dimension $D = 2$ and $D = 3$. The
cusp in the ${\mathcal E}_{\infty}(h)$ curve reflects a drastic
change of the multipartite entanglement around a critical point:
$h_{c}=3.1$ for $D = 2$ and $h_{c}=3.08$ for $D = 3$. Quantum Monte
Carlo simulation indicates a critical point at
$h_{\textrm{QMC}}\approx 3.044$.} \label{FIG3}
\end{figure}

{\it Simulation results.} We plot the geometric entanglement per
lattice site for quantum Ising model in a transverse field, with the
field strength $h$ as the control parameter, in Fig.~\ref{FIG3}.
There is a cusp at $h_{c}=3.10$ for the truncation dimension $D=2$
and $h_{c}=3.08$ for the truncation dimension $D=3$, respectively,
on the curve of the geometric entanglement as a function of the
transverse field $h$. On both sides of the cusp, it is continuous.
This implies that a continuous QPT occurs at $h_c$.

\begin{figure}
\includegraphics[width=0.42\textwidth]{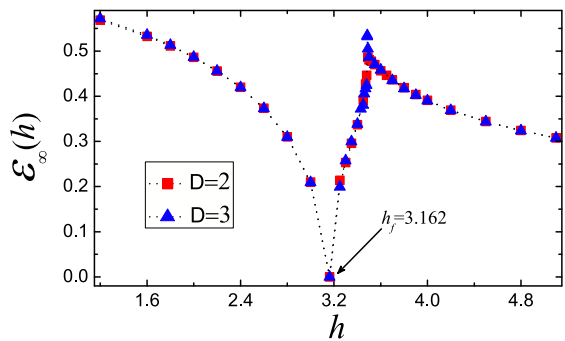}
\caption{(color online) The geometric entanglement per lattice site
${\mathcal E}_{\infty}(h)$ as a function of the applied external
magnetic field $h$ for quantum XYX model on an infinite-size square
lattice in two spatial dimensions, with $\Delta_{y} = 0.25$. The
data are presented for both $D = 2$ and $D = 3$. The factorizing
field, at which the geometric entanglement per lattice site
vanishes, is indicated by an arrow labeled by $h_{f}$ around 3.162.
Above the factorizing field $h_{f}$, a sharp increase of ${\mathcal
E}_{\infty}(h)$ reflects a rapid increase of the multipartite
entanglement around a critical point: $h_{c}=3.49$ for $D = 2$ and
$h_{c}=3.485$ for $D = 3$.} \label{FIG4}
\end{figure}

In Fig.~\ref{FIG4}, the geometric entanglement per lattice site for
quantum XYX model in an external magnetic field $h$, with
$\Delta_{y} = 0.25$, is plotted. Here, the external magnetic field
strength $h$ is chosen as the control parameter. A cusp occurs as
$h$ varies across a critical point $h_{c}$: $h_{c}=3.49$ for the
truncation dimension $D=2$ and $h_{c}=3.485$ for the truncation
dimension $D=3$, respectively. On both sides of the cusp, the
geometric entanglement per lattice site is continuous, which implies
that the model undergoes a continuous QPT at $h_{c}$.

Notice that, only a small shift is observed for $h_{c}$, with
increasing of the truncation dimension, for both models. This
indicates that the iPEPS algorithm captures many-body physics for
both models, with a small truncation dimension, as already observed
before~\cite{gv2d,zov,lsh}. Our simulation results are in a
quantitative agreement with Quantum Monte Carlo simulation for
quantum Ising model in a transverse field~\cite{hwj} and for quantum
XYX model in an external magnetic field $h$, with $\Delta_{y} =
0.25$~\cite{tros}. In addition, our simulation indicates that a
factorizing field $h_{f}$ occurs at $h_{f}=0$ and $h_{f}=3.162$,
respectively, for quantum Ising model in a transverse field and for
quantum XYX model in an external magnetic field $h$, with
$\Delta_{y} = 0.25$,  thus reproducing the exact results.  We stress
that a factorizing field $h_{f}$ is identified from  ${\mathcal
E}_{\infty}(h_{f}) =0$.

{\it Conclusions.} In this paper, we have demonstrated how to
efficiently compute the geometric entanglement per lattice site, by
optimizing over all the possible separable states, in the context of
the tensor network algorithm based on the iPEPS representation. The
geometric entanglement per lattice site, as a holistic measure of
the multipartite entanglement, serves as a universal marker to
locate critical points underlying quantum many-body lattice systems.
Our method is tested for both quantum Ising model in a transverse
magnetic field and an anisotropic spin $1/2$ anti-ferromagnetic XYX
model in an external magnetic field on an infinite-size square
lattice, succeeded in identifying both the critical points and
factorizing fields. We expect that, with the developments of
powerful tensor network algorithms, the geometric entanglement per
lattice site adds a new route to explore quantum criticality for
quantum many-body lattice systems in condensed matter physics.

{\it Acknowledgements.} The work is partially supported by the
National Natural Science Foundation of China (Grant No: 10874252).
HLW, QQS, and SHL are supported by the Fundamental Research Funds
for the Central Universities (Project No. CDJXS11102214) and by
Chongqing University Postgraduates' Science and Innovation Fund
(Project No.: 200911C1A0060322).

 
\end{document}